\documentclass{article}

\usepackage{arxiv}

\usepackage[utf8]{inputenc} 
\usepackage[T1]{fontenc}    
\usepackage{hyperref}       
\usepackage{url}            
\usepackage{booktabs}       
\usepackage{amsfonts}       
\usepackage{nicefrac}       
\usepackage{microtype}      
\usepackage{lipsum}
\usepackage{graphicx}
\usepackage{multirow}
\usepackage{authblk}
\usepackage{caption}
\usepackage{subcaption}
\usepackage{sistyle}
\SIthousandsep{,}

\usepackage{mathtools}

\usepackage{xcolor}

\title{Local Adaptation Improves Accuracy of Deep Learning Model for Automated X-Ray Thoracic Disease Detection : A Thai Study}

\author[1]{Isarun Chamveha}
\author[3]{Trongtum Tongdee}
\author[3]{Pairash Saiviroonporn}
\author[2]{Warasinee Chaisangmongkon}
\affil[1]{Perceptra Co., Ltd., Bangkok, Thailand}
\affil[ ]{\texttt{isarun@perceptra.tech}}
\affil[2]{Institute of Field Robotics, King Mongkut's University of Technology Thonburi, Bangkok, Thailand}
\affil[ ]{\texttt{warasinee.cha@mail.kmutt.ac.th}}
\affil[3]{Radiology Department, Faculty of Medicine Siriraj Hospital, Mahidol University, Bangkok, Thailand}
\affil[ ]{\texttt{\{trongtum, pairash.sai\}@gmail.com}}

\begin{document}
\maketitle

\begin{abstract}
Despite much promising research in the area of artificial intelligence for medical image diagnosis, there has been no large-scale validation study done in Thailand to confirm the accuracy and utility of such algorithms when applied to local datasets. Here we present a wide-reaching development and testing of a deep learning algorithm for automated thoracic disease detection, utilizing 421,859 local chest radiographs.  Our study shows that convolutional neural networks can achieve remarkable performance in detecting 13 common abnormality conditions on chest X-ray, and the incorporation of local images into the training set is key to the model's success. This paper presents a state-of-the-art model for CXR abnormality detection, reaching an average AUROC of 0.91. This model, if integrated to the workflow, can result in up to 55.6\% work reduction for medical practitioners in the CXR analysis process. Our work emphasizes the importance of investing in local research of medical diagnosis algorithms to ensure safe and efficient usage within the intended region.
\end{abstract}

\section{Introduction}

In modern clinical practice, chest X-ray, or CXR, is one of the most widely used methods in diagnosing abnormal conditions in the chest and nearby structures due to its noninvasive nature, low cost and high availability. Chest radiograph images can reveal abnormalities inside the lung parenchyma, mediastinum, rib cage, and heart, allowing physicians to determine the causes of various illnesses and monitor treatments for a variety of life-threatening diseases, such as pneumonia, tuberculosis, and cancer. In Western societies, CXR is performed 236 times on average on 1,000 patients, which accounts for $25\%$ of all diagnostic imaging procedures \cite{speets2006chest}. 

In Thailand and Southeast Asian neighboring countries, thoracic diseases constitute the leading causes of deaths for all age groups. Among the ten leading causes of deaths in SEA \cite{rao2013mortality}, five are thoracic abnormalities — Tuberculosis, lung cancer, Pneumonia, heart disease, and chronic obstructive pulmonary disease (COPD) — all of which rely on chest X-ray imaging as the primary method of diagnosis. As the region has the highest rate of Tuberculosis (TB) according to a World Health Organization (WHO) report \cite{whotbreport2019} and accounts for $44\%$ of TB cases worldwide, CXR has become prevalent in routine examinations and pre-employment screenings region-wide.

However, the analysis of chest radiographs requires years of training and experience. An X-ray projection produces a 2D image where anatomical structures can be obscured and abnormal features of various pathologies may be hard to differentiate. A study in the United Kingdom indicated that the difference in reading accuracy between a general practitioner and a radiologist specialist can be significant with concordance scores between $58\%$ to $74.6\%$ depending on years of medical practice \cite{satia2013assessing}. In Thailand, while there are tens of millions of chest X-ray images annually, there are only 1,277 radiologists in the whole country \cite{tmcstat2019}. It would be extremely difficult and cost-intensive to increase the number of qualified radiology specialists to make highly accurate diagnoses of CXR accessible to the general public at a sufficient rate to meet the ever-growing demands.

Recent advances in machine learning have introduced a wide variety of computer vision methods that can be used for medical image analysis. Deep learning is a growing trend in the field of computer-aided diagnosis for chest X-ray images, as several studies have demonstrated promising classification performance by convolutional neural networks \cite{rajpurkar2017chexnet, irvin2019chexpert, singh2018deep, guan2018diagnose}. With a wealth of open data released from medical centers around the world \cite{wang2017nihdataset, irvin2019chexpert, johnson2019mimic, bustos2019padchest}, any machine learning scientist can contribute to the advancement of AI algorithms for abnormality detection. Several proprietary algorithms have also shown amazing results comparing the diagnosis capabilities of radiologists and artificial intelligence \cite{hwang2019development, putha2018can}. However, some studies suggest that these impressive accuracies might not be transferable to unseen data sets, especially when algorithms trained with images from one population are transferred to make inference about another population \cite{sathitratanacheewin2018deep}. Overfitting and limited generalizability remain major issues in the applications of these algorithms in real-world settings.

In this work, we present the validation results of our deep learning algorithm for an automated detection of thoracic diseases in chest X-ray images from the Thai population. To our knowledge, this validation study incorporates the largest proprietary training dataset in Thailand from Siriraj Hospital, containing $421,859$ frontal chest X-ray images with verified radiologist reports. Our model was trained on a repository of over 1.1M CXR images collected from medical centers around the world. Our study aimed to achieve two goals. 

1. To validate deep learning algorithms in thoracic abnormality detection on a large and previously unseen pool of X-ray images from Thai population. This aim focuses on assessing the generalizability and practical utility of this approach for medical screening and diagnosis protocols. 

To this end, we present a state-of-the-art result from our automated CXR abnormality detection algorithm across 13 thoracic conditions in the Thai population. To our knowledge, this is the largest validation study done in the country. Our model has an average AUROC of 0.914 and generalizes well across unseen datasets. 

2. To investigate how well a deep-learning model can perform when trained on a large dataset of labeled images from other countries and tested on a dataset from the Thai population (i.e. Reference model). Moreover, we compare the Reference model with a model trained on data from Thai hospitals in addition to the data from other countries (i.e. Locally Adapted model) to determine if introduction of images from the local population will improve the model’s performance.

We demonstrate that the incorporation of local population data during training can result in drastic improvements in classification accuracies across conditions, with an average of 8.3\% improvement in prediction accuracy. We also quantified the difference in the lens of practical benefits; how much time can the AI save when incorporated into physicians’ workflow? A model that incorporates local data can result in a 46\% increase in screening efficiency, compared to a model trained exclusively with data outside the country. Our study has implications for the significance of supporting locally developed artificial intelligence solutions.
\section{Background}

\subsection{Chest Radiography}
Chest X-rays (CXR) are projection radiographs of the chest used to diagnose conditions affecting the chest, its contents, and nearby structures. Chest X-ray films are usually saved in Digital Imaging and Communications in Medicine (DICOM) format, which allows high-resolution images to be saved along with metadata such as patient name, gender and age. Radiologists diagnose the film and write up reports to describe visible conditions and store them along with chest X-ray films in the Picture Archiving and Communication System (PACS).

\subsection{Chest X-ray Abnormality Detection with Deep Learning}

In early years, various approaches using image processing techniques have been proposed to detect disease-specific features and identify abnormalities in chest X-ray images.

Dallal et al. \cite{dallal2017imgproc} proposed an approach to assess cardiomegaly condition in patients by measuring the cardiothoracic-ratio (CTR) value. Their approach searches a reference image repository for the nearest neighbor of the patient's sample X-ray image, and a SIFT flow algorithm was used to align and transform the lung boundary from the nearest neighbor image to the sample.

Udeshani et al. \cite{udeshani2011statistical} presents an algorithm for detecting lung cancer from chest X-ray images. After removing noise from the image, they segment regions suspected to have lung nodules by finding connected components from the image and calculate roundness index for each connected component. A neural network with one hidden layer is then applied on first-order features based on gray levels and second-order features generated from gray level co-occurrence matrix (GLCM) to verify if that region contains a nodule.

Various studies using image processing techniques for screening Tuberculosis from chest X-ray films were nicely presented in a detailed survey by Dallal et al. \cite{jaeger2013automatic}. While these approaches work well in the applications they are proposed for, they do not extend well to the detection of other diseases or to other datasets.

Convolutional Neural Networks (CNNs) has gained popularity in computer vision due to its ability to perform object detection tasks with great accuracy and generalizability.  Research has been conducted on medical abnormality screening from chest X-ray images using CNNs. Unlike image processing-based approaches, these research studies aim to detect every disease with one CNN model, making them more flexible to use and adapt to the detection of new diseases.

Among many neural network architectures used in medical image diagnosis, one notable line of work focuses on ResNet architecture. Proposed by He et al. \cite{he2016deep}, Resnet improved conventional CNNs by introducing residual blocks and skip connections that allow later layers to use activations from early layers, reducing the impact of vanishing gradients. This initial idea has been improved by the introduction of DenseNet by Huang et al. \cite{huang2017densely}, which adds dense blocks — a set of layers with dense connectivity. These connections allow later layers to fully utilize activation from earlier layers. By utilizing dense blocks, DenseNet achieved better results on ImageNet \cite{deng2009imagenet} dataset.

This family of neural networks has been widely utilized in X-ray image modeling. Wang et al. evaluated various CNN models on the ChestX-ray14 dataset and measured the accuracy \cite{wang2017nihdataset}. They found that the ResNet-50 model yields the best results on most chest conditions. Rajpurkar et al. used the DenseNet-121 model to detect abnormality in chest X-ray images \cite{rajpurkar2017chexnet}. Images are resized to $224$ by $224$ pixels before they are fed into the network. They reported that findings obtained by their deep learning model achieve comparable results to those made by radiologists and also improved the AUC score over previous approaches. The class activation mappings (CAMs) \cite{zhou2016learning} from their model applied on chest X-ray images also show regions similar to those radiologists use to diagnose chest X-ray images.

Irvin et al. experimented with DenseNet-121 trained on the CheXpert dataset from Stanford University \cite{irvin2019chexpert}. The model is trained with multi-view chest X-ray images with a resolution of $320$ by $320$ pixels. They compared the performances with three radiologists and reported that their approach achieves comparable performances to radiologists at each of their operating points. They also compared various approaches using uncertainty labels and concluded that treating such labels as positive yielded the best result on the detection of Atelectasis and treating them as a separate class yielded the best result on the detection of Cardiomegaly and was inconclusive on other classes.

Our work utilizes EfficientNet as proposed by Tan et al. \cite{tan2019efficientnet}, which improves upon traditional scaling of Convolutional Neural Networks (CNNs). By balancing network depth, width, and resolution, they achieve state-of-the-art results on the ImageNet \cite{deng2009imagenet} dataset compared with other models using the same parameter count. They also used the Swish activation function, which outperforms the traditional ReLU function on trained model accuracy. Additionally, they performed grid search on their scaling parameters and provided models of various sizes pre-trained on ImageNet, ranging from EfficientNet-B1, which has the smallest parameter count, to EfficientNet-B7, with the largest parameter count.

\section{Methodology}

\subsection{Dataset}

Chest X-ray data were collected over several years at Siriraj Hospital, Bangkok, Thailand. The dataset consists of $421,859$ frontal chest X-ray images along with their corresponding radiologist reports. All images were de-identified, and the usage of this dataset for this work was approved by Siriraj Institutional Review Board (SIRB). In this paper, we refer to this dataset as the Thai CXR dataset.  

To obtain ground-truth labels for machine learning model training, we used a modified version of the CheXpert labeler \cite{irvin2019chexpert} to generate labels with $13$ common radiographic chest abnormalities. Human readers verified the quality of a random subset of the extracted labels. Table \ref{table:abnormality_distribution} shows the distribution of each chest abnormality in our dataset.

\begin{table}[htbp]
\begin{center}
\caption{Number of each chest abnormality in the Thai Chest X-ray dataset}
\begin{tabular}{|c|c|c|c|c|}
\hline
\textbf{Pathology}&\textbf{Positive (\%)}&\textbf{Negative (\%)}&\textbf{Uncertain (\%)}&\textbf{No Mention (\%)}\\
\hline
Atelectasis&$6,404~(1.52)$&$492~(0.11)$& $8,241~(1.95)$&$406,722~(96.41)$\\
\hline
Cardiomegaly&$90,710~(21.50)$&$308,437~(73.11)$&$3,926~(0.93)$&$18,786~(4.45)$\\
\hline
Consolidation&$990~(0.23)$&$1,211~(0.29)$&$455~(0.11)$&$41,9203~(99.37)$\\
\hline
Edema&$3,933~(0.93)$&$2,087~(0.49)$&$2,458~(0.58)$&$413,381~(97.99)$\\
\hline
Lung Opacity&$23,472~(5.56)$&$200~(0.05)$&$1,689~(0.40)$&$396,498~(93.99)$\\
\hline
Pneumonia&$771~(0.18)$&$288~(0.07)$&$3,762~(0.89)$&$417,038~(98.86)$\\
\hline
Pleural Effusion&$14,753~(3.50)$&$147,378~(34.94)$&$10,487~(2.49)$&$249,241~(59.08)$\\
\hline
Pneumothorax&$632~(0.15)$&$13,616~(3.23)$&$133~(0.03)$&$407,478~(96.59)$\\
\hline
Pleural Other&$14,287~(3.39)$&$129~(0.03)$&$7107~(1.68)$&$400,336~(94.90)$\\
\hline
Hernia&$182~(0.04)$&$19~(0.00)$&$249~(0.06)$&$421,409~(99.89)$\\
\hline
Emphysema&$1,119~(0.27)$&$93~(0.02)$&$244~(0.06)$&$420,403~(99.65)$\\
\hline
Fracture&$2,088~(0.49)$&$785~(0.19)$&$271~(0.06)$&$418,715~(99.25)$\\
\hline
Support Devices&$20,616~(4.89)$&$430~(0.10)$&$388~(0.09)$&$400,425~(94.92)$\\
\hline
No Finding&$171577 (40.67)$&$0 (0.00)$&$0 (0.00)$&$250282 (59.33)$\\
\hline
\end{tabular}
\label{table:abnormality_distribution}
\end{center}
\end{table}

We applied a 60-20-20 train-validation-test split to divide the samples for machine learning.

\subsection{Automated Chest X-ray Abnormality Detection Algorithm}

We constructed two variants of deep learning models: the Reference model and the Locally Adapted model. In this paper, we use the word ‘model’ to refer to the specific configuration of architecture plus weights calculated from the training procedures.

The Reference model was trained on a repository of $705,184$ CXR images gathered from medical centers around the world, excluding Thailand. This dataset is referred to as Reference dataset in this paper. The dataset was split with 60-20-20 rule as in the Thai CXR dataset. Whenever applicable, we used the same labeler to extract class labels from radiologists' natural language reports. If the labels were supplied by the source, we accepted those labels as the ground truth and fitted them into the same 13-condition scheme.

The Locally Adapted model was trained on the Reference dataset combined with the Thai CXR dataset. All other parameters and training protocol were kept consistent.

Our abnormality detection model is created based on EfficientNet-B5 model from \cite{tan2019efficientnet} pretrained with ImageNet. This model has approximately 30 million parameters and has rated performance equal to models with $5.2$ times the number of parameters \cite{tan2019efficientnet}. We trained the EfficientNet-B5 model using Adam (Adaptive Moment Estimation) optimizer with a batch size of $16$ for $15$ epochs and an initial learning rate of $0.0001$. Images are resized to 456 by 456 pixels and are normalized with histogram equalization. 

To highlight the difference between the two models, we performed two tests. The first shows model performance using test data that was split from the Reference dataset (Reference in Table \ref{table:auc_comparison}). The second test shows performances of both models on unseen Thai CXR dataset (Thai-CXR in Table \ref{table:auc_comparison}). 
\section{Experiments and Results}

\subsection{Classification Performance Evaluation}

\begin{figure*}[htbp]

\begin{subfigure}{.21\textwidth}
  \centering
  \includegraphics[width=\linewidth]{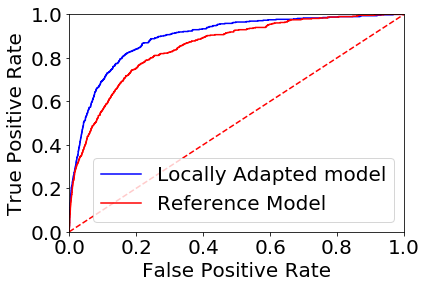}  
  \caption{Atelectasis}
\end{subfigure}
\begin{subfigure}{.21\textwidth}
  \centering
  \includegraphics[width=\linewidth]{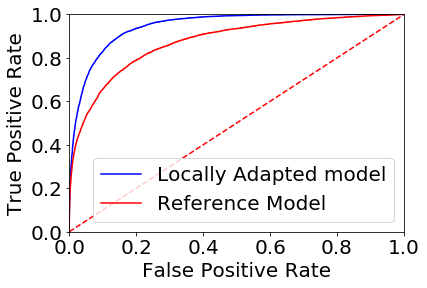}  
  \caption{Cardiomegaly}
\end{subfigure}
\begin{subfigure}{.21\textwidth}
  \centering
  \includegraphics[width=\linewidth]{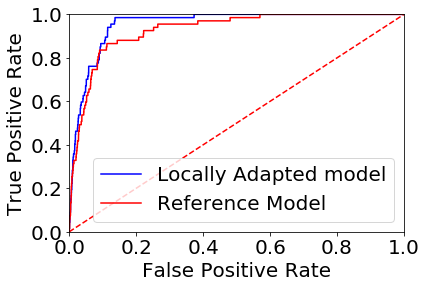}  
  \caption{Consolidation}
\end{subfigure}
\begin{subfigure}{.21\textwidth}
  \centering
  \includegraphics[width=\linewidth]{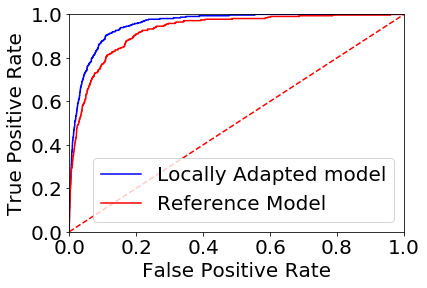}  
  \caption{Edema}
\end{subfigure}
\newline
\begin{subfigure}{.21\textwidth}
  \centering
  \includegraphics[width=\linewidth]{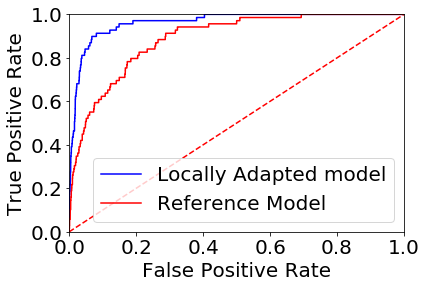}  
  \caption{Emphysema}
\end{subfigure}
\begin{subfigure}{.21\textwidth}
  \centering
  \includegraphics[width=\linewidth]{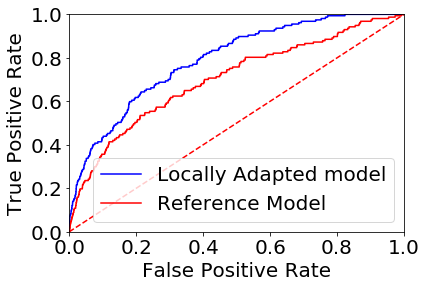}  
  \caption{Fracture}
\end{subfigure}
\begin{subfigure}{.21\textwidth}
  \centering
  \includegraphics[width=\linewidth]{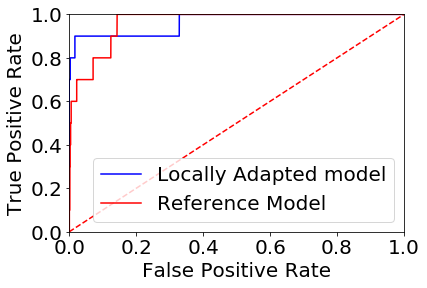}  
  \caption{Hernia}
\end{subfigure}
\begin{subfigure}{.21\textwidth}
  \centering
  \includegraphics[width=\linewidth]{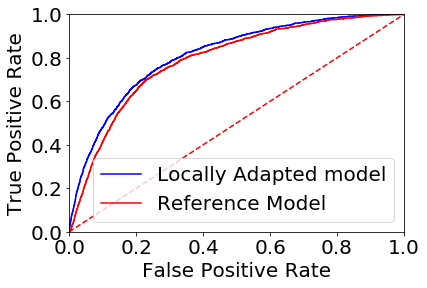}  
  \caption{Lung Opacity}
\end{subfigure}
\newline
\begin{subfigure}{.21\textwidth}
  \centering
  \includegraphics[width=\linewidth]{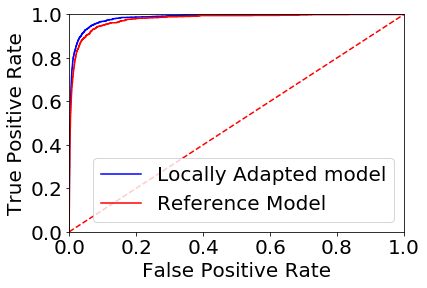}  
  \caption{Pleural Effusion}
\end{subfigure}
\begin{subfigure}{.21\textwidth}
  \centering
  \includegraphics[width=\linewidth]{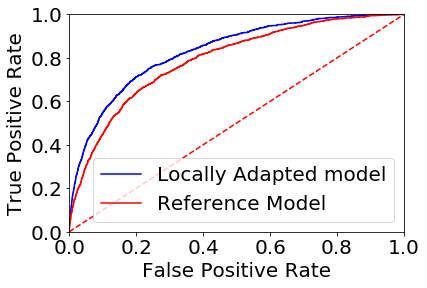}  
  \caption{Pleural Other}
\end{subfigure}
\begin{subfigure}{.21\textwidth}
  \centering
  \includegraphics[width=\linewidth]{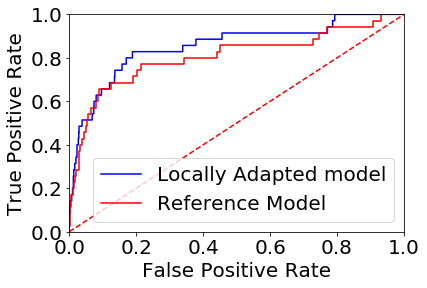}  
  \caption{Pneumonia}
\end{subfigure}
\begin{subfigure}{.21\textwidth}
  \centering
  \includegraphics[width=\linewidth]{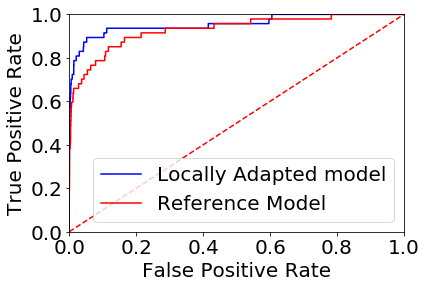}  
  \caption{Pneumothorax}
\end{subfigure}
\newline
\begin{subfigure}{.21\textwidth}
  \centering
  \includegraphics[width=\linewidth]{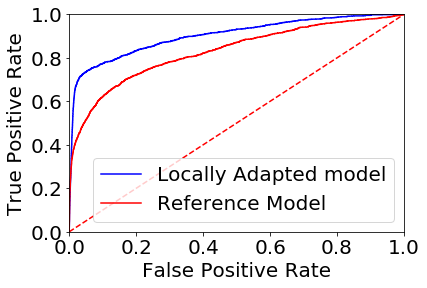}  
  \caption{Support Devices}
\end{subfigure}
\caption{Receiver Operating Characteristics (ROC) curve for each disease, comparing Locally Adapted model and Reference model on Thai Chest X-ray dataset. Figure is best viewed in color.}
\label{fig:roc_graph}
\end{figure*}

To compare classification performances of the Reference model and Locally Adapted model, we computed the area under receiver operating characteristic curve (AUROC) for each condition. Table \ref{table:auc_comparison} shows  the AUROC metrics for the two tests we conducted. Figure \ref{fig:roc_graph} shows the graph between true positive rate (TPR) and false positive rate (FPR) of our model compared with the Reference model.

Firstly, looking at the Reference model by itself, we see that it performs worse on data captured from a different domain (i.e. testing on Thai data vs Reference data), and the mean AUROC dropped significantly when tested with data from Thai population. Overall, the Reference model’s classification performance dropped from $0.880$ to $0.848$, or $3.8$\% drop when used in cross-population inference. This drop in accuracy means caution should be taken when using commercial software not specifically refined for Thai population.

On the contrary, the Locally Adapted model outperforms the Reference model in all conditions when tested on the Thai CXR data. The Locally Adapted model can achieve an average performance as high as $0.914$. Compared to AUROC of $0.848$ for the Reference model, this represents $8.3$\% improvement.
This can be explained by the training dataset for this model being more reflective of the test set, in terms of image quality, feature distribution, and label distribution. The improved labeler also potentially contributes to the improvement and we refer this to our future studies.
As a result, the model learns a better representation of the Thai chest X-ray data and yields better results on such data.
In other words, models trained with data from foreign medical institutions should be sufficiently adapted with data from Thai population before deploying in Thailand.


Despite being trained with a wide range of dataset, the Locally Adapted model still retained its performance on the Reference dataset at the AUROC of $0.894$, with a slight difference compared with the Reference model ($0.899$). This finding suggested that higher variability in the training data did not significantly degrade the performance and generalizability of the model.

\begin{table}[htbp]
\begin{center}
\caption{Area under ROC curve (AUROC) of each model displayed by each chest abnormality condition evaluated on the test subsets of the Reference dataset and Thai CXR dataset.}
\begin{tabular}{|c|c|c|c|c|}
\hline
\multirow{2}{*}{\textbf{Condition}}&\multicolumn{2}{c|}{\textbf{Locally Adapted Model}}&\multicolumn{2}{c|}{\textbf{Reference Model}}\\
\cline{2-5}
&\textbf{Reference}&\textbf{Thai-CXR}&\textbf{Reference}&\textbf{Thai-CXR}\\
\hline
Atelectasis&$0.811$&$\textbf{0.894}$&$\textbf{0.817}$&$0.881$\\
\hline
Cardiomegaly&$0.904$&$\textbf{0.946}$&$\textbf{0.912}$&$0.857$\\
\hline
Consolidation&$0.825$&$\textbf{0.954}$&$\textbf{0.831}$&$0.924$\\
\hline
Edema&$0.936$&$\textbf{0.956}$&$\textbf{0.939}$&$0.904$\\
\hline
Lung Opacity&$0.793$&$\textbf{0.813}$&$\textbf{0.802}$&$0.784$\\
\hline
Pneumonia&$0.837$&$\textbf{0.854}$&$\textbf{0.848}$&$0.743$\\
\hline
Pleural Effusion&$0.938$&$\textbf{0.985}$&$\textbf{0.940}$&$0.977$\\
\hline
Pneumothorax&$0.937$&$\textbf{0.955}$&$\textbf{0.942}$&$0.876$\\
\hline
Pleural Other&$0.885$&$\textbf{0.832}$&$\textbf{0.894}$&$0.809$\\
\hline
Hernia&$0.942$&$\textbf{0.965}$&$\textbf{0.969}$&$0.922$\\
\hline
Emphysema&$0.971$&$\textbf{0.963}$&$\textbf{0.982}$&$0.893$\\
\hline
Fracture&$\textbf{0.886}$&$\textbf{0.792}$&$0.885$&$0.697$\\
\hline
Support Devices&$0.968$&$\textbf{0.903}$&$\textbf{0.969}$&$0.803$\\
\hline
Mean&$0.894$&$\textbf{0.914}$&$\textbf{0.899}$&$0.848$\\
\hline
STD&$0.072$&$0.058$&$0.069$&$0.055$\\
\hline
\end{tabular}
\label{table:auc_comparison}
\end{center}
\end{table}



\section{Practical Utility Evaluation}
It is difficult to gauge the practical value of a machine learning model by just looking at AUROC measurements. How much impact can an $8.3$\% improvement in AUC really make for doctors and nurses who interact with the AI? Medical professionals are used to the metrics of sensitivity and specificity, which entails the probability of diagnostic errors. However, these measures are less informative since models' precision should be assessed across a range of thresholds, while sensitivity and specificity only measure probabilities at a single threshold.

Despite machine learning practitioners often showing that their algorithms outperform radiologists in medical image diagnosis, artificial intelligence applications in hospitals are still limited to the assistive role, instead of replacing doctors. 

However, it is useful to quantify the impact of artificial intelligence application under a realistic scenario of a chest X-ray analysis, we therefore developed a practical utility measure called Work Reduction. Our assumptions were based on an ideal case, where artificial intelligence is trusted to remove low-risk cases from the radiologists’ workload in order to allow them to focus on more difficult and high-risk cases.

In this ideal case, we set the algorithm to an operating point, at which it makes only 5\% false negative error. To hit that target, we would set a low threshold such that it yields a sensitivity value of $0.95$ for each disease. This ensures that only the normal cases with a high confidence of correct assignment are identified and that very few abnormal cases are misclassified as normal. At this conservative threshold, Work Reduction is the percentage of cases that the algorithm can de-prioritize (or classify as normal). The measurement is related to the specificity value at the high-sensitivity operating point as well as disease probability in a given dataset. 

Table \ref{table:sens_95_stats} shows the specificity and Work Reduction rate measured at the high-sensitivity operating point and tested on Thai CXR datasets. On average, our Locally Adapted model can reduce screening load by $55.9$\%. Compared to the Reference model at work reduction of $38.3$\%, our model can improve screening efficiency by $46.0$\%. 

\begin{figure}[htbp]
\centering
\includegraphics[width=0.5\linewidth]{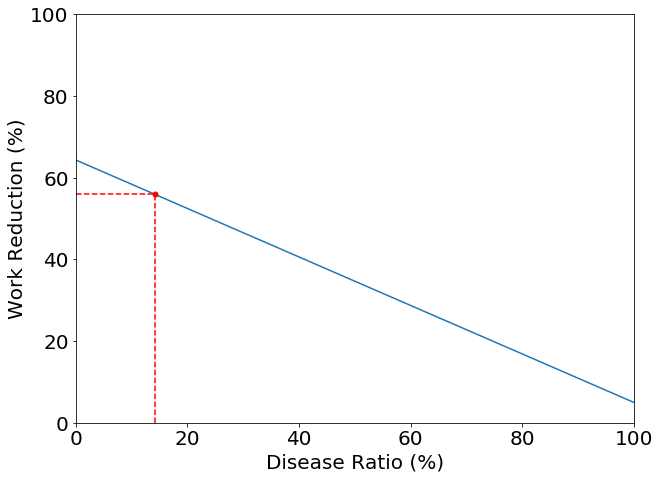}  
\caption{Relationship between average Work Reduction and disease ratio at the high-sensitivity operating point. Dotted lines show Work Reduction rate at the mean disease ratio of the dataset.}
\label{fig:wr_graph}
\end{figure}

In real-world usage, the Work Reduction rate also varies based on the disease ratio (i.e. the proportion of images in the dataset with a given abnormal condition).
Work Reduction rate in Table \ref{table:sens_95_stats} is calculated from the training dataset, which is taken from a tertiary-care hospital where moderate disease ratio is expected. In cases where most of the population are healthy, i.e., in a health check-ups, there will be a lower disease ratio and higher Work Reduction rate, whereas in an emergency room scenario, higher disease ratio and lower Work Reduction rate are expected.
Figure \ref{fig:wr_graph} shows the Work Reduction rate of the model at various disease ratios. The values calculated from the Locally Adapted model operating at the high-sensitivity point on the Thai Chest X-ray dataset are indicated by the red dot and corresponding dotted lines in the graph (with the average disease ratio of $14.2$\% across conditions). Our model can help with de-prioritizing cases most effectively in scenarios where a large proportion of samples are negative, such as in annual physical checkups. Readers should note that Work Reduction rates can vary in their specific use cases, depending on the probability distribution.

\begin{table}[htbp]
\begin{center}
\caption{Specificity and Work Reduction for each condition at high-sensitivity operating point (sensitivity fixed at $0.95$ for each chest abnormality).}
\begin{tabular}{|c|c|c|c|c|}
\hline
\multirow{2}{*}{\textbf{Condition}}&\multicolumn{2}{c|}{\textbf{Locally Adapted Model}}&\multicolumn{2}{c|}{\textbf{Reference Model}}\\
\cline{2-5}
&\textbf{Specificity}&\textbf{Work Reduction}&\textbf{Specificity}&\textbf{Work Reduction}\\
\hline
Atelectasis&$0.561$&$0.554$&$0.408$&$0.403$\\
\hline
Cardiomegaly&$0.771$&$0.618$&$0.434$&$0.353$\\
\hline
Consolidation&$0.876$&$0.874$&$0.735$&$0.734$\\
\hline
Edema&$0.821$&$0.815$&$0.718$&$0.713$\\
\hline
Lung Opacity&$0.347$&$0.333$&$0.306$&$0.294$\\
\hline
Pneumonia&$0.229$&$0.229$&$0.230$&$0.230$\\
\hline
Pleural Effusion&$0.932$&$0.902$&$0.899$&$0.871$\\
\hline
Pneumothorax&$0.584$&$0.584$&$0.567$&$0.567$\\
\hline
Pleural Other&$0.379$&$0.367$&$0.298$&$0.289$\\
\hline
Hernia&$0.983$&$0.983$&$0.876$&$0.875$\\
\hline
Emphysema&$0.851$&$0.849$&$0.584$&$0.583$\\
\hline
Fracture&$0.357$&$0.355$&$0.146$&$0.145$\\
\hline
Support Devices&$0.413$&$0.397$&$0.266$&$0.256$\\
\hline
Mean&$0.643$&$0.559$&$0.430$&$0.383$\\
\hline
STD&$0.212$&$0.173$&$0.175$&$0.174$\\
\hline
\end{tabular}
\label{table:sens_95_stats}
\end{center}
\end{table}


\section*{Conclusion}

We present a locally adapted state-of-the-art model for diagnosis of Thai chest X-ray images, coupled with analysis on the usability and utility of the artificial intelligence application in realistic clinical scenarios.
Even in high-sensitivity settings, we obtained a Work Reduction ratio of $56\%$, which translates to the amount of time saved for radiologists while keeping the false negative rate low. Apart from saving a significant amount of time for radiologists, the model can also alert radiologists on likely-abnormal cases they might have overlooked and thus provide an algorithmic "second opinion". By integrating this generalized machine learning model into chest X-ray assessment tools, we hope to alleviate the shortage of radiologists in Thailand and potentially in other countries as well.

Machine learning technology can lead to a radical improvement of operational efficiency in the process of screening, analysis, and diagnosis of medical images. Many studies around the world have shown promising results. It’s worth emphasizing that the road towards safe AI adoption in medicine requires proper validation tests, with sufficiently large and relevant datasets and with guidance from medical professionals. This is to ensure that the algorithm can generalize as well as follow medical common sense. Aside from best practices in modeling, users of artificial intelligence must be properly educated, so that they understand when they should trust the algorithm and when they should not. And finally, the end-user application may require multiple algorithms working together in a harmonious way to provide optimal performance reliability.

Future work shall explore whether this algorithm can achieve wide-range generalizability across samples from other Thai hospitals exhibiting high variability in image qualities, imaging configurations, and subjects' physical attributes.
Additional modern approaches to machine vision, such as attention-based mechanisms, super resolution, domain adaptation, explainable models, as well as label extraction techniques, are worth exploring to achieve even more robust results and optimal usability.

\bibliography{references}{}
\bibliographystyle{unsrt}

\end{document}